\documentclass[final,5p]{elsarticle}
\pdfoutput=1

\usepackage{amsmath,amssymb,amsthm}
\usepackage{amsfonts,mathrsfs}
\usepackage{lineno,hyperref}
\usepackage{tikz,tikzpagenodes}

\usepackage[mathscr]{eucal}

\usepackage{fourier} 
\usepackage[scaled=0.875]{helvet} 

\usepackage{microtype}

\usepackage{tikz}

\DeclareMathAlphabet\Euler{U}{eur}{m}{n}

\journal{arXiv}

\newcounter{appx} 

\newcommand{\phase}[1]{\medskip\noindent%
        \refstepcounter{appx}\textit{A.\arabic{appx}.\;}}

\newcommand{\nc}{\newcommand}

\theoremstyle{definition}

\theoremstyle{remark}

\nc{\hot}{\,\hat\otimes\,}
\nc{\ot}{\otimes}
\nc{\op}{\oplus}
\nc{\ol}{\overline}
\nc{\un}{\underline}

\nc{\mc}{\mathcal}
\nc{\ms}{\mathsf}
\nc{\mf}{\mathfrak}
\nc{\mb}{\mathbf}
\nc{\bb}{\mathbb}
\nc{\mr}{\mathrm}
\nc{\mt}{\mathscr}

\nc{\al}{\alpha}
\nc{\bet}{\beta}
\nc{\eps}{\epsilon}
\nc{\veps}{\varepsilon}
\nc{\del}{\delta}
\nc{\Del}{\Delta}
\nc{\ga}{\gamma}
\nc{\Ga}{\Gamma}
\nc{\ka}{\kappa}
\nc{\la}{\lambda}
\nc{\om}{\omega}
\nc{\si}{\sigma}
\nc{\Si}{\Sigma}
\nc{\Ups}{\upsilon}
\nc{\vphi}{\varphi}
\nc{\pa}{\partial}

\nc{\ii}{{\boldsymbol \imath}}
\nc{\ee}{{\mb e}}

\nc{\id}{\mathrm{id}}
\nc{\mfgr}{\mathrm{gr}}

\nc{\Ug}{U\mathfrak{g}}
\nc{\Ub}{U\mathfrak{b}}

\nc{\ud}{\underline}
\nc{\tl}{\tilde}
\nc{\wt}{\widetilde}
\nc{\wh}{\widehat}

\nc{\End}{\mathrm{End}}
\nc{\Ext}{\mathrm{Ext}}
\nc{\Hom}{\mathrm{Hom}}
\nc{\Ima}{\mathrm{Image}}
\nc{\Ind}{\mathrm{Ind}}
\nc{\Ker}{\mathrm{Ker}}
\nc{\betm}{\mathrm{RHom}}
\nc{\Sym}{\mathrm{Sym}}
\nc{\Str}{\mathrm{Str}}

\nc{\C}{\mathbb{C}}
\nc{\N}{\mathbb{N}}
\nc{\Z}{\mathbb{Z}}

\nc{\lan}{\langle}
\nc{\ran}{\rangle}

\nc{\mfg}{\mf{g}}
\nc{\mfa}{\mf{a}}

\nc{\mi}{{ \mathord{\begin{tikzpicture}[baseline=0.2ex, line width=.5, scale=0.1]
\draw[<-] (0,1) -- (1.5,1);
\end{tikzpicture}} }}

\nc{\lo}{{ \mathord{\begin{tikzpicture}[baseline=0.2ex, line width=.5, scale=0.1]
\draw[<-] (0,1) -- (1.1,1);
\draw[-] (1.1,0) -- (0,0);
\draw[domain=-90:90] plot ({1.1+.5*cos(\x)},{0.5+.5*sin(\x)});
\end{tikzpicture}} }}

\nc{\ua}{{\uparrow}}
\nc{\da}{{\downarrow}}
\nc{\uda}{{\ua\!\da}}

\nc{\cui}[1]{\ms{c}_{#1\ua}}
\nc{\cdi}[1]{\ms{c}_{#1\da}}
\nc{\cdui}[1]{\ms{c}^\dag_{#1\ua}}
\nc{\cddi}[1]{\ms{c}^\dag_{#1\da}}
\nc{\nmui}[1]{\ms{n}_{#1\ua}}
\nc{\nmdi}[1]{\ms{n}_{#1\da}}

\nc{\cu}{\ms{c}_{\ua}}
\nc{\cd}{\ms{c}_{\da}}
\nc{\cdu}{\ms{c}^\dag_{\ua}}
\nc{\cdd}{\ms{c}^\dag_{\da}}
\nc{\nmu}{\ms{n}_{\ua}}
\nc{\nmd}{\ms{n}_{\da}}

\nc{\xxx}{\text{\tiny$XXX$}}

\nc{\eqa}[1]{\begin{align}#1\end{align}}
\nc{\eqn}[1]{\begin{align*}#1\end{align*}}
\nc{\eq}[1]{\begin{equation}#1\end{equation}}
\nc{\spl}[1]{\begin{equation}\begin{aligned}#1\end{aligned}\end{equation}}
\nc\el{\nonumber\\}
\nc\nn{\nonumber}

\nc{\qu}{\quad}
\nc{\qq}{\qquad}

\nc{\ket}[1]{|#1\rangle}
\nc{\vac}{|0\rangle}

\nc{\ident}[1]{\vspace{0.1cm}\noindent{\bf #1}}

\nc{\Osc}{\Euler{Osc}}

\numberwithin{equation}{section}

\nc{\red}{\color{red}}
\nc{\blu}{\color{blue}}
\nc{\gre}{\color{green!50!black}}
\nc{\bur}{\color{red!50!black}}

\nc{\sfrac}[2]{{\textstyle\frac{#1}{#2}}}
\nc{\half}{\sfrac{1}{2}}
\nc{\ihalf}{\sfrac{\ii}{2}}
\nc{\quarter}{\sfrac{1}{4}}

\nc{\alg}[1]{\mathfrak{#1}}
\nc{\rep}{\rho}

\begin{document}


\begin{frontmatter}

\title{How to fold a spin chain: \\ Integrable boundaries of the Heisenberg XXX and Inozemtsev hyperbolic models}

\author{Alejandro De La Rosa Gomez}
\ead{alrg500@york.ac.uk}

\author{Niall MacKay}
\ead{nm15@york.ac.uk}

\author{Vidas Regelskis}
\ead{vr509@york.ac.uk}

\address{Department of Mathematics, University of York \\ York, YO10 5DD, United Kingdom}

\begin{abstract}
We present a general method of folding an integrable spin chain, defined on a line, to obtain an integrable open spin chain, defined on a half-line. We illustrate our method through two fundamental models with $\mf{sl}_2$ Lie algebra symmetry: the Heisenberg XXX and the Inozemtsev hyperbolic spin chains. We obtain new long-range boundary Hamiltonians and demonstrate that they exhibit Yangian symmetries, thus ensuring integrability of the models we obtain. The method presented provides a ``bottom-up'' approach for constructing integrable boundaries and can be applied to any spin chain model.
\end{abstract}


\end{frontmatter}


\section{Introduction}

The standard picture of boundary integrability of 1+1D spin chains and quantum field theories in the quantum inverse scattering method, due to Sklyanin \cite{Sklyanin}, contains an implicit idea of ``folding''. It begins with bulk and boundary Yang-Baxter equations, and their associated $R$- and $K$-matrices, and uses the latter to construct a boundary transfer matrix from its bulk parent. This process contains an implied folding of the infinite line (or chain) back on itself to create a half-line, and thereby a boundary-integrable model on this half-line. This folding is only rarely made explicit in the physics literature \cite{DM} but has been studied in the context of Temperley-Lieb and blob algebras \cite{MGP,AR1,AR2}.

However, this process can be difficult to implement in explicit cases. It does not begin with the Hamiltonian but rather extracts it, together with other conserved quantities and symmetries, from the transfer matrix. If we instead begin with a bulk Hamiltonian and wish to discover integrable boundaries and their symmetries, a different, ``bottom-up'' procedure is needed. This procedure, which we refer to as ``folding'', is a map denoted by $f$ (and $\overline f$ in the double-row case -- see below) which sends the spin operators and conserved charges of a model defined on an infinite chain to those defined on a semi-infinite chain. The purpose of this letter is to detail this procedure and apply it first to a classic and then to an overarching new case.

We begin with the elementary examples of the classic Heisenberg XXX spin chain and a ``double-row'' model of two XXX chains uncoupled except by the boundary. The latter is motivated by a similar structure which emerges in gauge/string (``AdS/CFT'') duality \cite{MR} and also serves as a toy model for the open Hubbard chain with an achiral boundary \cite{UK, Gomez}. We then go on to construct integrable boundaries for the Inozemtsev long-range infinite spin chain \cite{Ino1} and its doubling. In each case we emphasize the Yangian symmetry of the bulk model, and from it derive a twisted Yangian symmetry of the model with an integrable boundary.

The XXX and Inozemtsev spin chains are the natural choices to work with. The former is the most famous, prototypical spin chain in the physics literature. It allows us to check that the results obtained in this paper are in agreement with well-known ones, and also introduces the reader to our procedure through a relatively simple example.

The Inozemtsev chain, by contrast, is less well-known, but may be the more fundamental. All famous $\mf{sl}_2$ spin chains are limiting cases of it (see Section \ref{sec:4}). It also possesses striking thermodynamic properties of its own \cite{Klabbers}. But most importantly for modern fundamental physics, it appears in the context of AdS/CFT -- in particular, the expression for the dilatation operator of $\mc{N}=4$ SYM in the planar limit coincides with its conserved charges up to three loops \cite{SYM1, SYM2}.

The main motivation for our folding procedure is that, to construct integrable boundaries for long range spin chains like Inozemtsev's, one cannot use the boundary Yang Baxter equation in the usual way and must instead rely on Dunkl operators \cite{BPS, XW}. This is where our bottom-up approach becomes useful: starting with a long range Hamiltonian defined on the infinite line, our folding procedure allows us to systematically construct integrable boundaries without the use of a monodromy matrix.

This letter is organized as follows. In Section \ref{sec:2} we set up the chain and explain its folding. In Section \ref{sec:3} we study folding of the infinite XXX spin chain. The methods obtained are then used in Section \ref{sec:4} to fold an Inozemtsev hyperbolic spin chain. Section \ref{sec:5} contains concluding remarks and a discussion of relevant open questions. The appendix contains details of the Yangian algebras studied in this letter. 

Most of the results presented were computed using the {\tt Wolfram Mathematica} computer algebra system. For readers' convenience we have detailed explicitly some of the computations that explain the folding of the Hamiltonian and Yangian operators.


\section{Setting up the spin chain} \label{sec:2}


\paragraph{Lattice}

Fix $L \in \N$ and consider a one-dimensional lattice with $2L$ sites that can be occupied by spin-$1$ particles. Each lattice site is identified with a two-dimensional vector space $V_i\cong \C^{2}$ spanned by vectors
\eq{
V_i = {\rm span}_\C \{\, \ket{\ua}_i ,\; \ket{\da}_i \}, \label{V_i}
}
where $-L< i\le L$ is the index of the site in the lattice. The entire lattice is the $2L$-fold tensor product $V := \bigotimes_{-L < i \le L} V_i$.

To describe dynamics of such a lattice, we employ Pauli matrices $\si^x_i$, $\si^y_i$, $\si^z_i$ and the identity matrix $\si^0_i$ that satisfy the usual (anti-)commutation relations
\eq{
\{\si^a_i,\si^b_j\} = 2 \delta_{ij} \del_{ab} \si^0_i, \qu[\si^a_i,\si^b_j] = 2 \ii \delta_{ij} \veps_{abc} \si^c_i \label{Pauli}
}
where $\ii = \sqrt{-1}$ denotes the imaginary unit, $a,b,c\in\{x,y,z\}$ and $\veps_{abc}$ is the Levi-Civita symbol normalized so that $\veps_{xyz}=1$ and we have used the Einstein summation rule of the repeated indices.
Then, upon introducing $\si^\pm_i = \half (\si_i^x \pm \ii \si^y_i)$, we require that, for $i\ne j$,
\spl{
\si^+_i \ket{\da}_i &= \ket{\ua}_i , &&
\si^z_i \ket{\ua}_i = \ket{\ua}_i , &&
\si^a_i \ket{\ua}_j = \ket{\ua}_j\, \si^a_i ,
\\
\si^-_i \ket{\ua}_i &= \ket{\da}_i , &&
\si^z_i \ket{\da}_i = -\ket{\da}_i , &&
\si^a_i \ket{\da}_j = \ket{\da}_j\, \si^a_i . \label{Pauli:action}
}

Matrices $\si^a_i$ provide a unitary representation of the universal envelope $U(\mf{sl}_2)$ of the $\mf{sl}_2$ Lie algebra
\eq{
\rho_L \;:\; x^\pm \mapsto \sum_{-L< i\le L}\si^\pm_i, \qu h \mapsto \sum_{-L< i\le L} \si^z_i, \label{rho}
}
where $x^\pm$, $h$ are the standard generators of the $\mf{sl}_2$ Lie algebra satisfying $[x^+,x^-]=h$ and $[h,x^\pm]=\pm2x^\pm$. The map \eqref{rho} together with \eqref{Pauli:action} turns the vector space $V$ into a left $U(\mf{sl}_2)$-module.


\paragraph{Folding}

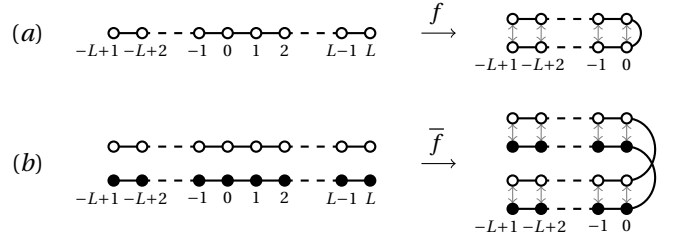
\begin{figure}

\eqn{
(a)\qu&\begin{tikzpicture}[scale=75/100,baseline=-0.35em]
\draw[thick] (0,0) -- (.5,0);
\draw[thick] (1.5,0) -- (3,0);
\draw[thick] (4,0) -- (4.5,0);
\draw[thick,dashed] (.5,0) -- (1.5,0);
\draw[thick,dashed] (3,0) -- (4,0);
\draw[thick,fill=white] (0,0) circle (.09) node[below=1pt]{\scriptsize $-L\!+\!1\qq$};
\draw[thick,fill=white] (0.5,0) circle (.09) node[below=1pt]{\scriptsize $\;\;-L\!+\!2$};
\draw[thick,fill=white] (1.5,0) circle (.09) node[below=1pt]{\scriptsize $-1\;$};
\draw[thick,fill=white] (2,0) circle (.09) node[below=1pt]{\scriptsize $0$};
\draw[thick,fill=white] (2.5,0) circle (.09) node[below=1pt]{\scriptsize $1$};
\draw[thick,fill=white] (3,0) circle (.09) node[below=1pt]{\scriptsize $2$};
\draw[thick,fill=white] (4,0) circle (.09) node[below=1pt]{\scriptsize $L\!-\!1$};
\draw[thick,fill=white] (4.5,0) circle (.09) node[below=1pt]{\scriptsize $L$};
\draw[thick] (7,.25) -- (7.5,.25);
\draw[thick] (8.5,.25) -- (9,.25);
\draw[thick] (7,-.25) -- (7.5,-.25);
\draw[thick] (8.5,-.25) -- (9,-.25);
\draw[thick,dashed] (7.5,.25) -- (8.5,.25);
\draw[thick,dashed] (7.5,-.25) -- (8.5,-.25);
\draw[thick,domain=-90:90] plot ({9+.25*cos(\x)},{.25*sin(\x)});
\draw[<->,gray] (7,.15) -- (7,-.15);
\draw[<->,gray] (7.5,.15) -- (7.5,-.15);
\draw[<->,gray] (8.5,.15) -- (8.5,-.15);
\draw[<->,gray] (9,.15) -- (9,-.15);
\draw[thick,fill=white] (7,.25) circle (.09);
\draw[thick,fill=white] (7.5,.25) circle (.09);
\draw[thick,fill=white] (8.5,.25) circle (.09);
\draw[thick,fill=white] (9,.25) circle (.09);
\draw[thick,fill=white] (7,-.25) circle (.09) node[below=1pt]{\scriptsize $-L\!+\!1\qq$};
\draw[thick,fill=white] (7.5,-.25) circle (.09) node[below=1pt]{\scriptsize $\;-L\!+\!2$};
\draw[thick,fill=white] (8.5,-.25) circle (.09) node[below=1pt]{\scriptsize $-1\;$};
\draw[thick,fill=white] (9,-.25) circle (.09) node[below=1pt]{\scriptsize $0$};
\draw[->] (5.4,0) -- (6,0) node[above]{\small$\;f\qq$};
\end{tikzpicture}
\\[1em]
(b)\qu&\begin{tikzpicture}[scale=75/100,baseline=.9em]
\draw[thick] (0,.8) -- (.5,.8);
\draw[thick] (1.5,.8) -- (3,.8);
\draw[thick] (4,.8) -- (4.5,.8);
\draw[thick,dashed] (.5,.8) -- (1.5,.8);
\draw[thick,dashed] (3,.8) -- (4,.8);
\draw[thick,fill=white] (0,.8) circle (.09);
\draw[thick,fill=white] (0.5,.8) circle (.09);
\draw[thick,fill=white] (1.5,.8) circle (.09);
\draw[thick,fill=white] (2,.8) circle (.09);
\draw[thick,fill=white] (2.5,.8) circle (.09);
\draw[thick,fill=white] (3,.8) circle (.09);
\draw[thick,fill=white] (4,.8) circle (.09);
\draw[thick,fill=white] (4.5,.8) circle (.09);
\draw[thick] (0,.2) -- (.5,.2);
\draw[thick] (1.5,.2) -- (3,.2);
\draw[thick] (4,.2) -- (4.5,.2);
\draw[thick,dashed] (.5,.2) -- (1.5,.2);
\draw[thick,dashed] (3,.2) -- (4,.2);
\draw[thick,fill=black] (0,.2) circle (.09) node[below=1pt]{\scriptsize $-L\!+\!1\qq$};
\draw[thick,fill=black] (0.5,.2) circle (.09) node[below=1pt]{\scriptsize $\;\;-L\!+\!2$};
\draw[thick,fill=black] (1.5,.2) circle (.09) node[below=1pt]{\scriptsize $-1\;$};
\draw[thick,fill=black] (2,.2) circle (.09) node[below=1pt]{\scriptsize $0$};
\draw[thick,fill=black] (2.5,.2) circle (.09) node[below=1pt]{\scriptsize $1$};
\draw[thick,fill=black] (3,.2) circle (.09) node[below=1pt]{\scriptsize $2$};
\draw[thick,fill=black] (4,.2) circle (.09) node[below=1pt]{\scriptsize $L\!-\!1$};
\draw[thick,fill=black] (4.5,.2) circle (.09) node[below=1pt]{\scriptsize $L$};
\draw[thick] (7,1.3) -- (7.5,1.3);
\draw[thick] (8.5,1.3) -- (9,1.3);
\draw[thick] (7,.8) -- (7.5,.8);
\draw[thick] (8.5,.8) -- (9,.8);
\draw[thick,dashed] (7.5,1.3) -- (8.5,1.3);
\draw[thick,dashed] (7.5,.8) -- (8.5,.8);
\draw[thick] (7,.2) -- (7.5,.2);
\draw[thick] (8.5,.2) -- (9,.2);
\draw[thick] (7,-.3) -- (7.5,-.3);
\draw[thick] (8.5,-.3) -- (9,-.3);
\draw[thick,dashed] (7.5,0.2) -- (8.5,.2);
\draw[thick,dashed] (7.5,-.3) -- (8.5,-.3);
\draw[thick,domain=-90:90] plot ({9+.5*cos(\x)},{0.75+.55*sin(\x)});
\draw[thick,domain=-90:90] plot ({9+.5*cos(\x)},{0.25+.55*sin(\x)});
\draw[<->,gray] (7,1.2) -- (7,.9);
\draw[<->,gray] (7.5,1.2) -- (7.5,.9);
\draw[<->,gray] (8.5,1.2) -- (8.5,.9);
\draw[<->,gray] (9,1.2) -- (9,.9);
\draw[<->,gray] (7,.1) -- (7,-.2);
\draw[<->,gray] (7.5,.1) -- (7.5,-.2);
\draw[<->,gray] (8.5,.1) -- (8.5,-.2);
\draw[<->,gray] (9,.1) -- (9,-.2);
\draw[thick,fill=white] (7,1.3) circle (.09);
\draw[thick,fill=white] (7.5,1.3) circle (.09);
\draw[thick,fill=white] (8.5,1.3) circle (.09);
\draw[thick,fill=white] (9,1.3) circle (.09);
\draw[thick,fill=black] (7,.8) circle (.09);
\draw[thick,fill=black] (7.5,.8) circle (.09);
\draw[thick,fill=black] (8.5,.8) circle (.09);
\draw[thick,fill=black] (9,.8) circle (.09);
\draw[thick,fill=white] (7,.2) circle (.09);
\draw[thick,fill=white] (7.5,.2) circle (.09);
\draw[thick,fill=white] (8.5,.2) circle (.09);
\draw[thick,fill=white] (9,.2) circle (.09);
\draw[thick,fill=black] (7,-.3) circle (.09) node[below=1pt]{\scriptsize $-L\!+\!1\qq$};
\draw[thick,fill=black] (7.5,-.3) circle (.09) node[below=1pt]{\scriptsize $\;\;-L\!+\!2$};
\draw[thick,fill=black] (8.5,-.3) circle (.09) node[below=1pt]{\scriptsize $-1$};
\draw[thick,fill=black] (9,-.3) circle (.09) node[below=1pt]{\scriptsize $0$};
\draw[->] (5.4,0.5) -- (6,0.5) node[above]{\small$\;\overline f\qq$};
\end{tikzpicture}
}
\caption{Folding: (a) Single-row lattice, (b) Double-row lattice.} \label{F:1}
\end{figure}

We fold the lattice by identifying sites labelled by indices $1\le i \le L$ with those labelled by $1-i$ as shown in Figure \ref{F:1} (a). We say that the lattice is folded over a link.

Let us explain how the folding acts on the matrices~$\si^a_i$. Recall that
\begin{gather*}
\si^\pm_i \si^z_i = \mp\si^\pm_i , \qu
\si^z_i \si^\pm_i = \pm\si^\pm_i , \qu \si^z_i \si^z_i = \si^0_i, \\
\si^\pm_i\si^\pm_i=0 , \qu \si^\pm_i \si^\mp_i = \half (\si^0_i \pm \si^z_i) ,
\end{gather*}
which imply that any polynomial in $\si^a_i$ can be written as a linear combinations of monomials
\eq{
\prod_{-L< i \le L} \si^{a_i}_{i} \qu \text{with} \qu a_i \in \{\pm,z,0\}, \label{monoms}
}
or in other words the monomials \eqref{monoms} provide a vector space basis of $\Sigma_L = \lan\,\si^a_i:a\in\{\pm,z\},\,-L< i\le L\,\ran$ over the field of complex numbers $\C$. Note that elements of $\Sigma$ are also elements of $\End V$; the element $\prod_{-L< i \le L} \si^0_i$ is the identity map.

Set $\Sigma^-_L = \lan\,\si^a_i:a\in\{\pm,z\},\, -L< i\le 0\,\ran $. We define the multiplicative folding $f : \Sigma_L \to \Sigma^-_L$ acting on monomials \eqref{monoms} by
\eq{
f : \prod_{-L< i \le L} \si^{a_i}_i \mapsto \prod_{-L< i \le 0} k^{a_{i}a_{1-i}}\, \si^{a_{i}}_{i} \si^{a_{1-i}}_{i} , \label{fold}
}
where $k^{a_{i}a_{1-i}} \in \C$ are model-depending folding constants that will be specified in the examples studied below.


\paragraph{Double-row lattice}

We will also consider folding of a double-row lattice, shown in Figure \ref{F:1} (b), which is a $2L$-fold tensor product of two copies of spaces $V_i$ additionally decorated by $\circ$~and~$\bullet$, that is $\overline V = \bigotimes_{-L< i \le L} (V_{i,\circ} \ot V_{i,\bullet})$.
The dynamics of such a lattice is described by decorated Pauli matrices ({\it c.f.}~\eqref{Pauli})
\[
{}[\si^a_{i\al},\si^b_{j\bet}] = 2 \ii \del_{ij} \del_{\al\bet} \veps_{abc} \si^c_{i\al}, \;\;\;
\{\si^a_{i\al},\si^b_{j\bet}\} = 2 \del_{ij} \del_{ab} \del_{\al\bet}\si^0_{i\al} ,
\]
where $\al,\bet\in\{\circ,\bullet\}$.

Any polynomial in matrices $\si^a_{i\al}$ can be written as a linear combinations of monomials ({\it c.f.}~\eqref{monoms})
\eq{
\prod_{-L< i \le L} \si^{a_i}_{i,\circ}\, \si^{b_i}_{i,\bullet} \qu \text{with} \qu a_i,b_i \in \{\pm,z,0\}, \label{monoms2}
}
that provide a vector space basis of the double-row analogue of $\Si_L$, namely $\overline\Sigma_L = \lan\,\si^a_{i\al}:a\in\{\pm,z\},\,1\le i\le L,\,\al\in\{\circ,\bullet\}\,\ran$.

Let $\overline{\Si}{}^{-}_L$ be a double-row analogue of $\Si^{-}_L$. We define the multiplicative folding $\overline{f} : \overline{\Si}_L \to \overline{\Si}{}^{-}_L$ acting on monomials \eqref{monoms2} by ({\it c.f.}~\eqref{fold})
\spl{
\overline f : \prod_{-L < i \le L} \si^{a_i}_{i,\circ}\, \si^{b_i}_{i,\bullet} \mapsto \prod_{-L< i \le 0} k^{a_{i}b_{1-i}} k^{b_{i}a_{1-i}}\, \si^{a_{i}}_{i,\circ} \si^{b_{1-i}}_{i,\circ} \si^{b_{i}}_{i,\bullet} \si^{a_{1-i}}_{i,\bullet}  \label{bfold}
}
where $k^{a_{i}b_{1-i}},k^{b_{i}a_{1-i}} \in \C$ are model-depending folding constants that will be specified in the examples studied below. Note that folding constants are labelled by indices $\pm,z,0$ only. We treat both rows, labelled by $\circ$ and $\bullet$, on an equal footing.


\section{Heisenberg XXX spin chain} \label{sec:3}


\paragraph{Infinite chain}

It is well known that the Hamiltonian of the Heisenberg XXX spin chain
\eq{
\mt{H}_{\xxx} = -\la \sum_{-L< i\le L} \big( \si^+_i \si^-_{i+1} + \si^-_i \si^+_{i+1} + \half \si^z_i \si^z_{i+1} \big) \label{Hxxx}
}
commutes with the Lie operators $\mt{E}_0^\pm = \rho_L(x^\pm)$ and $\mt{E}_0^z = \rho_L(h)$. We say that the Hamiltonian $\mt{H}_{\xxx}$ exhibits a $U(\mf{sl}_2)$ Lie algebra symmetry.

When the chain is infinitely long, {\em i.e.\ }$L\to\infty$, the Hamiltonian $\mt{H}_{\xxx}$ additionally exhibits a Yangian symmetry. More precisely, it commutes, up to the terms at infinity, with the operators
\spl{
\mt{E}_1^{\prime\pm} &= \pm \tfrac{\la}{2} \sum_{i<j} \si^{\pm}_i\si^{z}_j , &
\mt{E}_1^{\prime z} &= \la\sum_{i<j} \si^{+}_i\si^{-}_j ,
\\
\mt{E}_1^{\prime\prime\pm} &= \mp \tfrac{\la}{2} \sum_{i<j} \si^{z}_i\si^{\pm}_j , &
\mt{E}_1^{\prime\prime z} &= -\la\sum_{i<j} \si^{-}_i\si^{+}_j , \label{half-Yang1}
}
which, combined to
\eq{
\mt{E}_1^{\pm} = \mt{E}_1^{\prime\pm} + \mt{E}_1^{\prime\prime\pm} , \qq
\mt{E}_1^{z} = \mt{E}_1^{\prime z} + \mt{E}_1^{\prime\prime z} , \label{Yang1}
}
satisfy the defining relations of the Yangian $\mc{Y}(\mf{sl}_2)$ \cite{Bernard},  see \ref{A1}. We will say that $\mt{E}^\pm_1$ and $\mt{E}^z_1$ are Yangian operators.
Note that the sum $\sum_{i<j}$ in \eqref{half-Yang1} is understood as $\sum_{-\infty <i<j <\infty}$. We will use a similar notation in further sections; for example, $\sum_{i\le0}$ will be understood as $\sum_{-\infty <i\le 0}$.


\paragraph{Magnetic boundary}

Let us now focus on a semi-infinite spin chain with a boundary magnetic field described by \cite{Gaudin}
\eq{
\mt{H}_{\xxx}^\mu =\mt{H}_{\xxx}^\mi + \mu\sigma^z_0 , \label{HxxxM}
}
where $\mt{H}_{\xxx}^\mi$ denotes the XXX spin chain Hamiltonian with sites labelled from $-\infty$ to 0 (we will use the notation $(\;)^\mi$ for all operators restricted to a semi-infinite chain) and $\mu\si^z_0$ is the boundary term with $\mu \in \C$ being the strength of a boundary magnetic field.

The presence of the boundary term in \eqref{HxxxM} breaks the $\mc{Y}(\mf{sl}_2)$ Yangian symmetry down to the $\mc{Y}^+(\mf{sl}_2)$ twisted Yangian. In particular, the Hamiltonian $\mt{H}_\xxx^\mu$ commutes with $(\mt{E}^z_0)^\mi$ and, up to the terms at infinity, with Yangian operators $\mt{X}^\pm$ defined by \cite{NepDoi,Doikou2}
\eq{
\mt{X}^{\pm} = (\mt{E}^{\pm}_1)^{\mi} \pm \tfrac{\la}{2} (\mt{E}_0^{\pm})^\mi (\mt{E}^z_0)^\mi + \tfrac{\la}{2}\big(1 \mp \tfrac{\la}{\mu}\big)(\mt{E}^{\pm}_0)^\mi , \label{Yang2}
}
that are elements in $\Si^\mi_\infty$ and satisfy the defining relations of $\mc{Y}^+(\mf{sl}_2)$, see \ref{A2}. It is worth noting that operators
\spl{
\mt{X}^{\prime\pm} &= (\mt{E}^{\prime\pm}_1)^{\mi} \mp \tfrac{\la^2}{4\mu}(\mt{E}^{\pm}_0)^\mi , \\
\mt{X}^{\prime\prime\pm} &= (\mt{E}^{\prime\prime\pm}_1)^{\mi} \pm \tfrac{\la}{2} (\mt{E}_0^{\pm})^\mi (\mt{E}^z_0)^\mi + \tfrac{\la}{2}\big(1 \mp \tfrac{\la}{2\mu}\big)(\mt{E}^{\pm}_0)^\mi , \label{half-Yang2}
}
satisfying $\mt{X}^\pm = \mt{X}^{\prime\pm} + \mt{X}^{\prime\prime\pm}$,
are also symmetries of $\mt{H}^\mu_\xxx$. They can be views as analogues of the symmetries \eqref{half-Yang1} of $\mt{H}_\xxx$.

Our goal is to demonstrate the method of obtaining the Hamiltonian $\mt{H}_\xxx^\mu$ from $\mt{H}_\xxx$ and Yangian operators \eqref{Yang2} from those in \eqref{Yang1} by employing the folding \eqref{fold}.
The first step is to impose the following constraints on the folding constants:
\eq{
k^{\pm0}=- k^{0\pm}=k^{z0}= k^{0z} =1, \qu k^{\pm z}= k^{z \pm}, \label{k1}
}
which ensure that Lie symmetries of $\mt{H}_\xxx$ are projected to those of $\mt{H}^\mu_\xxx$. Recall that $U(\mf{sl}_2)$, as a vector space, is linearly spanned by the monomials $f^l h^m e^n$ with $l,m,n\in \Z_{\ge0}$. Thus we must make sure that any monomial $(\mt{E}^-_0)^l (\mt{E}^z_0)^m (\mt{E}^+_0)^n$ for any $l,m,n\in \Z_{\ge0}$, each being a symmetry of $\mt{H}_\xxx$, is folded into a symmetry of $\mt{H}^\mu_\xxx$, which exhibits a $U(\mf{gl}_1)\subset U(\mf{sl}_2)$ symmetry only. The first constraint in \eqref{k1} yields
\eqn{
f(\mt{E}^\pm_0) &= f(\sum_{i} \si^\pm_i) = (k^{\pm0}+k^{0\pm}) \sum_{i\le 0} \si^\pm_i = 0, \\
f(\mt{E}^z_0) &= f(\sum_{i} \si^\pm_i) = (k^{z0}+k^{0z}) \sum_{i\le 0} \si^z_i = 2 \sum_{i\le 0} \si^z_i = 2(\mt{E}^z_0)^\mi,
}
while second constraint in \eqref{k1} additionally ensures that any monomial $(\mt{E}^-_0)^l(\mt{E}^z_0)^m(\mt{E}^+_0)^n$ is folded into a symmetry of $\mt{H}^\mu_\xxx$. In particular, for any $l,m,n \in \Z_{\ge0}$, we have that
\[
f((\mt{E}^-_0)^l(\mt{E}^z_0)^m(\mt{E}^+_0)^n) = \del_{ln} \sum_{0\le r\le l+m} c_r\,((\mt{E}^z_0)^\mi)^r
\]
for some $c_r \in \C$.
Note that $k^{\pm\pm}$ do not play a role in the folding, since $\si^\pm_i\si^\pm_i=0$. We also set $k^{zz}=1$, so that $f(\rho_L(h^l))=f(\rho_L(h^m)) f(\rho_L(h^n))$ for any $l,m,n\in\Z_{+}$ satisfying $l=m+n$. (We will comment on this property in Section \ref{sec:5}.)

Next, using \eqref{k1} and splitting the sum $\sum_{i}$ into three terms as $\sum_{i} = \sum_{i<0} + \del_{i0} + \sum_{i>0}$, we fold the Hamiltonian $\mt{H}_\xxx$ of the infinite chain:
\eqa{
\hspace{.3cm} & \hspace{-.3cm} f(\mt{H}_{\xxx}) = \nn
\\
& = -\la \Bigg( \sum_{i<0} \left( k^{+0}k^{-0}(\si^+_i \si^-_{i+1} + \si^-_i \si^+_{i+1}) + \half(k^{z0})^2\si^z_i \si^z_{i+1} \right)  \nn
\\
& \qq + k^{+-}\si^+_0 \si^-_{0} + k^{-+}\si^-_0 \si^+_0 + \half k^{zz} \si^z_0 \si^z_0 \nn
\\
& \qq + \sum_{i>0} \left( k^{0+}k^{0-}(\si^+_{1-i} \si^-_{-i} + \si^-_{1-i} \si^+_{-i}) + \half(k^{0z})^2\si^z_{1-i} \si^z_{-i} \right) \Bigg) \nn
\\
& = 2 \mt{H}^\mi_{\xxx} - \tfrac{\la}{2}\left((k^{+-}-k^{-+})\,\si^z_0+(1+k^{+-}+k^{-+})\right). \label{fH1}
}
Choosing $k^{-+}-k^{+-}=\frac{4 \mu}{\lambda}$ we have that $f(\mt{H}_{\xxx}) = 2 \mt{H}_{\xxx}^\mu$ up to a constant term.

In order to fold the Yangian operators \eqref{Yang1} we first split the sum $\sum_{i<j}$ into four terms
\eq{
\textstyle \sum_{i<j\le 0} + \del_{i+j\ne1} \sum_{i\le 0<j} + \del_{i+j=1}\sum_{i\le 0<j} + \sum_{0<i<j}. \label{split}
}
By doing so for \eqref{Yang1} and folding each sum individually we find
\eqa{
f(\mt{E}_1^{z}) &= \la \Bigg(\sum_{i<j\le 0}\big(k^{+0}k^{-0} \si^+_i \si^-_j-k^{-0}k^{+0} \si^-_i \si^+_j\big)
\el
& \qq\qu + k^{0-} \big((\mt{E}_0^+)^\mi (\mt{E}_0^-)^\mi-\sum_{i\leq 0}\si_i^+\si_i^-\big)
\el
& \qq\qu - k^{0+}\big((\mt{E}_0^-)^\mi( \mt{E}_0^+)^\mi -\sum_{i\leq 0}\si_i^-\si_i^+\big)
\el
& \qq\qu + \sum_{i\leq 0}\big( k^{+-} \si_i^+\si_i^- - k^{-+} \si_i^-\si_i^+ \big)
\el
& \qq\qu +\sum_{0<i<j}\big(k^{0+}k^{0-} \si^+_{1-i} \si^-_{1-j}-k^{0-}k^{0+} \si^-_{1-i} \si^+_{1-j}\big)
\Bigg)
\el
&= \tfrac{\la}{2} L (k^{-+}-k^{+-}) - \tfrac{\la}{2}(k^{+-}+k^{-+})(\mt{E}_0^z)^\mi \label{fE1z}
}
which commutes with $f(\mt{H}_\xxx)$, and
\eqa{
f(\mt{E}_1^{\pm}) &= \pm\tfrac{\la}{2} \Bigg(\sum_{i<j\le 0} \big(k^{\pm 0}k^{z0} \si^{\pm}_i \si^z_j-k^{z0}k^{\pm 0} \si^z_i \si^{\pm}_j\big)
\el
& \qq\qu + k^{0z} \big((\mt{E}_0^{\pm})^\mi (\mt{E}_0^z)^\mi-\sum_{i\leq 0}\si_i^{\pm}\si_i^z\big)
\el
& \qq\qu - k^{0 \pm}\big((\mt{E}_0^z)^\mi( \mt{E}_0^{\pm})^\mi -\sum_{i\leq 0}\si_i^z\si_i^{\pm}\big)
\el
& \qq\qu + \sum_{i\leq 0}\big( k^{\pm z} \si_i^{\pm}\si_i^z - k^{z \pm} \si_i^z\si_i^{\pm} \big)
\el
& \qq\qu +\sum_{0<i<j}\big(k^{0\pm}k^{0z} \si^{\pm}_{1-i} \si^z_{1-j}-k^{0z}k^{0\pm} \si^z_{1-i} \si^{\pm}_{1-j}\big)
\Bigg)
\el
&=2\big( (\mt{E}_1^{\pm})^\mi \pm \tfrac{\la}{2} (\mt{E}^{\pm}_0)^\mi (\mt{E}^z_0)^\mi + \tfrac{\la}{2}(1-k^{z \pm})(\mt{E}_0^{\pm})^\mi \big) , \label{fE1pm}
}
which commute with $f(\mt{H}_\xxx)$, up to the terms at infinity, only~if
\eq{
k^{z\pm} = \mp\frac{4}{k^{+-} - k^{-+}} = \pm \frac{\la}{\mu} , \label{k2}
}
in which case we obtain $f(\mt{E}_1^{\pm})=2 \mt{X}^{\pm}$, as expected. We also have that $f(\mt{E}_1^{\prime\pm})=2 \mt{X}^{\prime\pm}$ and $f(\mt{E}_1^{\prime\prime\pm})=2 \mt{X}^{\prime\prime\pm}$, so that the summetries \eqref{half-Yang1} of $\mt{H}_\xxx$ are folded into the symmetries \eqref{half-Yang2} of $\mt{H}^\mi_\xxx$. Thus we have demonstrated that with a suitable choice of the folding constants, which were deduced from the symmetry arguments, the Hamiltonian $\mt{H}_\xxx$ of the infinite spin chain and its symmetries can be folded into the Hamiltonian $\mt{H}^\mu_\xxx$ of a semi-infinite spin chain with a magnetic boundary and its symmetries.

In the remaining parts of this section we will demonstrate how to obtain the semi-infinite spin chain with an open boundary and a semi-infinite double-row spin chain with a diagonal boundary. The obtained results will then be used in Section \ref{sec:4} to obtain the corresponding boundary models for the Inozemtsev hyperbolic spin chain.


\paragraph{Open boundary}

Setting the boundary magnetic field strength to $\mu=0$ in \eqref{HxxxM} we obtain a semi-infinite spin chain with an open boundary, namely
\eq{
\mt{H}^{0}_{\xxx}=\mt{H}^\mi_{\xxx}. \label{Hxxx0}
}
This Hamiltonian exhibits a $U(\mf{sl}_2)$ symmetry by commuting with operators $(\mt{E}^\pm_0)^\mi$ and $(\mt{E}^z_0)^\mi$, but does not commute with those in \eqref{Yang1} viewed as elements in $\Si^\mi_{\infty}$. However, upon defining higher-order Yangian operators
\eq{
\mt{E}^\pm_2 = \pm \tfrac{1}{2}[\mt{E}^z_1,\mt{E}^\pm_1], \qu  \mt{E}^z_2 =[\mt{E}^+_1,\mt{E}^-_1] \label{level2}
}
the Hamiltonian $\mt{H}^0_\xxx$ commutes, up to the terms at infinity, with the operators~\footnote{It seems likely that these symmetries were observed before; however, we have been unable to locate them in the literature available to us.}
\spl{
\mt{G}^z &=(\mt{E}_2^z)^\mi - \la \left((\mt{E}^+_1)^\mi (\mt{E}_0^-)^\mi -(\mt{E}^+_0)^\mi (\mt{E}^-_1)^\mi \right) - \tfrac{\la^2}{4} (\mt{E}^z_0)^\mi , \\
\mt{G}^\pm &=(\mt{E}_2^\pm)^\mi \mp \tfrac{\la}{2} \left((\mt{E}_1^z)^\mi (\mt{E}_0^\pm)^\mi -(\mt{E}^z_0)^\mi (\mt{E}^\pm_1)^\mi \right) - \tfrac{\la^2}{4} (\mt{E}^\pm_0)^\mi , \;\, \label{YangG}
}
instead, that, together with $(\mt{E}^\pm_0)^\mi$ and $(\mt{E}^z_0)^\mi$, satisfy the defining relations of the $\mc{Y}^-(\mf{sl}_2)$ twisted Yangian, see \ref{A3}.

We now use the folding to obtain the Hamiltonian $\mt{H}^0_\xxx$ and its symmetries $(\mt{E}^a_0)^\mi$ and $\mt{G}^a$ with $a\in\{\pm,z\}$. Since the model exhibits a $U(\mf{sl}_2)$ symmetry it is natural to choose $k^{ab}=1$ for all $a,b \in \{\pm,z,0\}$. This gives
\eqn{
f(\mt{E}^\pm_0) &= f(\sum_{i} \si^\pm_i) = (k^{\pm0}+k^{0\pm}) \sum_{i\le 0} \si^\pm_i = 2 \sum_{i\leq 0} \si^{\pm}_i = 2(\mt{E}^\pm_0)^\mi,
\\
f(\mt{E}^z_0) &= f(\sum_{i} \si^\pm_i) = (k^{z0}+k^{0z}) \sum_{i\le 0} \si^z_i = 2 \sum_{i\le 0} \si^z_i = 2(\mt{E}^z_0)^\mi .
}
In a similar way one can check that with this choice of folding constants any monomial $(\mt{E}^-_0)^l (\mt{E}^z_0)^m (\mt{E}^+_0)^n$ for any $l,m,n\in \Z_{\ge0}$ is folded into a symmetry of $\mt{H}^0_\xxx$.

By folding the Hamiltonian $\mt{H}_\xxx$ we get ({\it c.f.}~\eqref{fH1})
\eqa{
f(\mt{H}_{\xxx}) &= 2\mt{H}^\mi_\xxx - \la \big(\si^+_0 \si^-_{0} + \si^-_0 \si^+_0 + \half\si^z_0 \si^z_0 \big) ,
}
which equals to $2\mt{H}^\mi_\xxx-\tfrac32\la$ and thus agrees with \eqref{Hxxx0} up to the constant term.

Folding Yangian operators \eqref{Yang1} we find $f(\mt{E}^a_1)=-\la(\mt{E}^a_0)^\mi$, which can be easily deduced from \eqref{fE1z} and \eqref{fE1pm}, and is in agreement with the fact that $\mc{Y}^-(\mf{sl}_2)$ twisted Yangian has elements of even grading only. Finally, we want to obtain the operators \eqref{YangG} using the folding. By folding the higher-order Yangian operators \eqref{level2} we obtain symmetries of $\mt{H}^0_\xxx$: folded operators $f(\mt{E}^\pm_2)$ and $f(\mt{E}^z_2)$ commute with $\mt{H}^0_\xxx$, up to the terms at infinity. However, the obtained operators do not coincide to those in \eqref{YangG}. It turns out that we need to fold the following operators
\eqa{
\wt{\mt{E}}^+_2 &= \mt{E}^+_2 + \tfrac13 \big([\mt{E}^{\prime z}_1, \mt{E}^{\prime\prime +}_1] + [\mt{E}^{\prime\prime z}_1, \mt{E}^{\prime +}_1] \big) + \tfrac{\la^2}{3} \big( \mt{E}^+_0 \mt{E}^-_0\mt{E}^+_0 - \tfrac94 \mt{E}_0^+ \big) ,
\el	
\wt{\mt{E}}^-_2 &= \mt{E}^-_2 - \tfrac13 \big([\mt{E}^{\prime z}_1, \mt{E}^{\prime -}_1] + [\mt{E}^{\prime\prime z}_1, \mt{E}^{\prime\prime -}_1] \big) + \tfrac{\la^2}{3} \big( \mt{E}^-_0 \mt{E}^+_0\mt{E}^-_0 - \tfrac94 \mt{E}_0^- \big) ,
\el
\wt{\mt{E}}^z_2 &= \mt{E}^z_2 + \tfrac23 \big( [\mt{E}^{\prime +}_1, \mt{E}^{\prime -}_1] + [\mt{E}^{\prime\prime +}_1, \mt{E}^{\prime\prime -}_1] \big) + \tfrac{\la^2}{6} \big( (\mt{E}^z_0)^3 - \tfrac72 \mt{E}_0^z \big) , \label{Goodlev2}
}
instead. The additional terms in the expressions above are symmetries of $\mt{H}_\xxx$ and are tailored in such a way that the operators $\wt{\mt{E}}^\pm_2$ and $\wt{\mt{E}}^z_2$ fold precisely to those in \eqref{YangG}, up to an overall scalar factor,
\eq{
f(\wt{\mt{E}}^\pm_2) = \tfrac83 \mt{G}^\pm, \qu f(\wt{\mt{E}}^z_2) = \tfrac83 \mt{G}^z . \label{foldEG}
}
The explicit form of computations in \eqref{foldEG} is very similar to those presented in \eqref{fE1z} and \eqref{fE1pm}, only the expressions are much more lengthy, thus we have not written them out explicitly. It will be shown in Section \ref{sec:4} that long-range analogues of $\wt{\mt{E}}^\pm_2$ and $\wt{\mt{E}}^z_2$ fold into Yangian symmetries of the long-range open boundary model.

As the final remark, we note that the open boundary model also exhibits a number of additional symmetries that are obtained by folding quadratic combinations of the operators in \eqref{half-Yang1}.


\paragraph{Double-row chain with a diagonal boundary}

Our third example of an integrable boundary model arises in the context of the double-row model consisting of two uncoupled XXX spin chains. The Hamiltonian of the latter is given by 
\eq{
\mt{H}_{\xxx}^{\circ\bullet} = -\la \sum_{\al=\circ,\bullet}\sum_{L<i \leq L} \big( \si^+_{i\al} \si^-_{i+1,\al} + \si^-_{i\al} \si^+_{i+1,\al} + \half \si^z_{i\al} \si^z_{i+1,\al} \big). \label{HxxxD}
}
In the $L\to\infty$ limit it exhibits a $\mc{Y}_\circ(\mf{sl}_2) \ot \mc{Y}_\bullet(\mf{sl}_2) \cong \mc{Y}( \mf{so}_4)$ symmetry expressed in terms of the Lie operators $\mt{E}^a_{0\al}$ and $\mt{E}^a_{1\al}$ with $a\in\{\pm,z\}$ and $\al\in\{\circ,\bullet\}$ that are the natural analogues of $\mt{E}^a_{0}$ and $\mt{E}^a_{1}$ for the double-row model.

Introduce linear combinations of Lie operators
\eq{
\mt{A}^{a}_n = \mt{E}^{a}_{n,\circ}+\mt{E}^{a}_{n,\bullet}, \qu
\mt{B}^{a}_n = \mt{E}^{a}_{n,\circ}-\mt{E}^{a}_{n,\bullet} \label{so4}
}
for all $a\in\{\pm,z\}$ and $n\in\{0,1\}$. Then the semi-infinite double-row Hamiltonian with a diagonal boundary
\eq{
\mt{H}_{\xxx}^{\Delta} = (\mt{H}^{\circ\bullet}_{\xxx})^\mi - \la\big( \si_{0,\circ}^+\si_{0,\bullet}^- + \si_{0,\circ}^-\si_{0,\bullet}^+ + \half \si_{0,\circ}^z \si_{0,\bullet}^z\big)\label{HxxxD2}
}
exhibits a diagonal $U(\mf{sl}^\Delta_2)\subset U(\mf{sl}^\circ_2) \ot U(\mf{sl}^\bullet_2)$ symmetry; it commutes with operators $(\mt{A}^{a}_0)^\mi$ only. The boundary term couples the two, otherwise uncoupled, spin-chains and can be viewed as a permutation operator; a similar boundary in the context of the Hubbard model was studied in \cite{Gomez}. Moreover, the double-row model with a diagonal boundary can also be viewed as an infinite spin-chain with a defect located at the middle of the chain. 

In the infinite limit, when $L\to\infty$, the Hamiltonian $\mt{H}^{\Delta}_\xxx$ additionally commutes, up to the terms at infinity, with the Yangian operators
\spl{
\mt{Y}^\pm &=(\mt{B}_1^{\pm})^\mi \pm \tfrac{\la}{4}((\mt{B}^{\pm}_0)^\mi (\mt{A}^z_0)^\mi-(\mt{A}_0^{\pm})^\mi (\mt{B}^z_0)^\mi),
\\
\mt{Y}^z &=(\mt{B}_1^{z})^\mi - \tfrac{\la}{2}((\mt{B}^{+}_0)^\mi (\mt{A}^-_0)^\mi-(\mt{A}_0^+)^\mi (\mt{B}^-_0)^\mi) \label{Yang3}
}
that, together with $(\mt{A}^a_0)^\mi$, satisfy the defining relations of the $\mc{Y}^{{\Delta}}(\mf{sl}_2)$ twisted Yangian, see \ref{A4}.

As for the open boundary case, we set $k^{ab}=1$ for all $a,b \in \{\pm,z,0\}$. Then a straightforward computation shows that the folding $\overline f$ acts on the operators defined in \eqref{so4} by
\[
\overline f(\mt{A}^a_0)=2(\mt{A}^a_0)^\mi, \quad \overline f(\mt{B}^a_0)=0 ,
\]
and on the Hamiltonian \eqref{HxxxD} by
\eqn{
\overline f &(\mt{H}^{\circ\bullet}_{\xxx}) = \\
&= -\la \sum_{\al=\circ,\bullet} \sum_{i<0} \big(\si_{i,\al}^+\si_{i+1,\al}^- + \si_{i,\al}^-\si_{i+1,\al}^+ +\tfrac{1}{2} \si_{i,\al}^z \si_{i+1,\al}^z \big)
\\ &\qu -2\la\big(\si_{0,\circ}^+\si_{0,\bullet}^- + \si_{0,\circ}^-\si_{0,\bullet}^+ +\tfrac{1}{2} \si_{0,\circ}^z \si_{0,\bullet}^z \big)
\\
&\qu- \la \sum_{\al=\circ,\bullet} \sum_{i< 0} \big(\si_{1-i,\al}^+\si_{-i,\al}^- + \si_{1-i,\al}^-\si_{-i,\al}^+ +\tfrac{1}{2} \si_{1-i,\al}^z \si_{-i,\al}^z \big)
\\
&=2(\mt{H}_{\xxx}^{\circ\bullet})^\mi
- 2\la\big(\si_{0,\circ}^+\si_{0,\bullet}^- + \si_{0,\circ}^-\si_{0,\bullet}^+ +\tfrac{1}{2} \si_{0,\circ}^z \si_{0,\bullet}^z \big)
}
thus exactly reproducing \eqref{HxxxD2}.

Applying the folding to the operators $\mt{B}^a_1$ we recover the ones defined in \eqref{Yang3}. In particular
\eqn{
\overline f(\mt{B}_1^{\pm}) &= \pm\tfrac{\la}{2} \Big(\sum_{i<j \leq 0} (\si^{\pm}_{i,\circ} \si^z_{j,\circ}-\si^z_{i,\circ} \si^{\pm}_{j,\circ}-\si^{\pm}_{i,\bullet} \si^z_{j,\bullet}+\si^{z}_{i,\bullet} \si^{\pm}_{j,\bullet})
\\
& \qq\qu +2\big((\mt{E}_{0,\circ}^{\pm})^\mi(\mt{E}_{0,\bullet}^z)^\mi-(\mt{E}_{0,\bullet}^{\pm})^\mi(\mt{E}_{0,\circ}^z)^\mi
\\
& \qq\qu -\sum_{i \leq 0}(\si_{i,\bullet}^z \si_{i,\circ}^{\pm}-\si_{i,\circ}^z \si_{i,\bullet}^{\pm})\big)
\\
& \qq\qu +2\sum_{i \leq 0}(\si_{i,\circ}^{\pm} \si_{i,\bullet}^z-\si_{i,\bullet}^{\pm} \si_{i,\circ}^z)
\\
&\qq\qu +\sum_{\al=\circ,\bullet} \sum_{0<i<j} (\si^{\pm}_{1-i,\al} \si^z_{1-j,\al}-\si^z_{1-i,\al} \si^{\pm}_{1-j,\al})\Big)
\\
&=2 \mt{B}^{\pm}_1 \pm \la \big((\mt{E}_{0,\circ}^{\pm})^\mi(\mt{E}_{0,\bullet}^z)^\mi-(\mt{E}_{0,\bullet}^{\pm})^\mi(\mt{E}_{0,\circ}^z)^\mi\big)=2\mt{Y}^{\pm}
 }
and
\eqn{
\overline f(\mt{B}_1^{z}) &= \la \Big(\sum_{i<j \leq 0} (\si^{+}_{i,\circ} \si^-_{j,\circ}-\si^-_{i,\circ} \si^{+}_{j,\circ}-\si^{+}_{i,\bullet} \si^-_{j,\bullet}+\si^{-}_{i,\bullet} \si^+_{j,\bullet})
\\
& \qq\qu +2\big((\mt{E}_{0,\circ}^{+})^\mi(\mt{E}_{0,\bullet}^-)^\mi-(\mt{E}_{0,\bullet}^{+})^\mi(\mt{E}_{0,\circ}^-)^\mi
\\
& \qq\qu-\sum_{i \leq 0}(\si_{i,\bullet}^- \si_{i,\circ}^{+}-\si_{i,\circ}^- \si_{i,\bullet}^{+})\big)
\\
& \qq\qu + 2\sum_{i \leq 0} (\si_{i,\circ}^{+} \si_{i,\bullet}^--\si_{i,\bullet}^{+} \si_{i,\circ}^-)
\\
&\qq\qu +\sum_{\al=\circ,\bullet} \sum_{0<i<j} (\si^{+}_{1-i,\al} \si^-_{1-j,\al}-\si^-_{1-i,\al} \si^{+}_{1-j,\al})\Big)
\\
&=2 \mt{B}^{z}_1 - 2\la \big((\mt{E}_{0,\circ}^{+})^\mi(\mt{E}_{0,\bullet}^-)^\mi-(\mt{E}_{0,\bullet}^{+})^\mi(\mt{E}_{0,\circ}^-)^\mi\big)=2\mt{Y}^{z} .
}
Repeating the same steps for $\mt{A}^a_1$ we find
\eq{
\overline f(\mt{A}^\pm_1) = \overline f(\mt{A}^z_1) = 0,
}
as expected. We conclude this section with a remark that the double-row model also exhibits additional symmetries that are natural analogues of those in \eqref{half-Yang1}.


\section{Inozemtsev hyperbolic spin chain} \label{sec:4}


\paragraph{Infinite chain}

The Inozemtsev elliptic spin chain is the long-range analogue of the XXX spin chain with Hamiltonian defined by
\eqa{
\mt{H}_{\ka} = -\tfrac{\la}{2} \sum_{\substack{-L<i,j\leq L \\ i \neq j}} \wp_L(i-j) \big( \si^+_i \si^-_{j} + \si^-_i \si^+_{j} + \tfrac12\si^z_i \si^z_{j} \big),
\label{HamIno}}
where $\wp_L$ is the Weierstra\ss\ elliptic function with periods $L$ and $\ii\pi/\ka$ for $\ka\in\bb{R}_{\ge0}$. This model exhibits a $U(\mf{sl}_2)$ symmetry identical to its nearest neighbour counterpart. By taking an appropriate limit of the parameter $\ka$ and the length $L$ this model specializes to the Haldane-Shastry, XXX and Inozemtsev hyperbolic (also called ``infinite'') spin chain. To see this, we need to rescale the hopping matrix of~\eqref{HamIno}
 \eq{
\widehat{\wp}_L(z):=\frac{\sinh^2(\ka)}{\ka^2}\Big(\wp_L(z)+\frac{2 \ka}{\ii \pi}\zeta_L\Big(\frac{\ii \pi}{2 \ka}\Big)\Big) ,
}
where $\zeta_L$ is the Weierstra\ss\ $\zeta$-function with quasiperiods $L$ and $ \ii \pi / \ka$. In the $\ka\to\infty$ limit one has
\eq{
\lim_{\ka\to\infty} \widehat{\wp}_L(z) =\delta_{z \ \mathrm{mod} \ L,1} ,
}
which recovers the XXX spin chain \cite{Ino2}. One can also take the $\ka\to 0$ limit to obtain the Haldane-Shastry hopping matrix \cite{HS}:
\eq{
\lim_{\ka\to 0} \widehat{\wp}_L(z) = \frac{\pi^2}{L^2 \sin^2(\pi z/L)} .
}
The limit we are interested in is when the length of the chain becomes infinite. In this case,
\eq{
p_{z} := \lim_{L\to\infty} \widehat{\wp}_L(z) = \frac{\sinh^2(\ka)}{\sinh^2(\ka z)}
\label{HM}}
and the $U(\mf{sl_2})$ symmetry can be enhanced to the $\mc{Y}(\mf{sl}_2)$ Yangian by introducing the operators
\eq{
\mt{E}^{\pm}_{\ka,1}=\pm \frac{\la}{2}\sum_{i,j} w_{i-j}\si_i^{\pm} \si_j^z, \qq
\mt{E}^z_{\ka,1}= \la\sum_{i,j}w_{i-j}\si_i^- \si_j^+, \label{YangIno}
}
where $w_z=-\coth(\kappa z)$ when $z \neq 0$ and $w_{0}=0$. These operators commute with the Hamiltonian, up to the terms at infinity, and satisfy the defining relations of the $\mc{Y}(\mf{sl}_2)$ Yangian.

The Hamiltonian also commutes, up to the terms at infinity, with operators $\mt{E}^{\prime a}_{\ka,1}$ and $\mt{E}^{\prime\prime a}_{\ka,1}$ defined analogously to $\mt{E}^{a}_{\ka,1}$,
$$
w'_{z} = \frac{\ee^{-\ka z}}{\ee^{-\ka z} - \ee^{\ka z}}  \qu\text{and}\qu
w''_{z} = \frac{\ee^{\ka z}}{\ee^{-\ka z} - \ee^{\ka z}} 
$$
respectively. We also set $w'_0=w''_0=0$, so that $w_z = w'_z + w''_z$. These operators are the long-range analogues of those in \eqref{half-Yang1}. In particular,
$$
\lim_{\ka\to\infty} w'_z = \del_{z<0} , \qq
\lim_{\ka\to\infty} w''_z = -\del_{z>0}.
$$

In the remaining part of this section we will use foldings $f$ and $\overline f$ studied in Section \ref{sec:3} to obtain integrable long-range boundary Hamiltonians and operators that commute with them.
From now on, $\mt{H}_{\ka}$ will denote
\eqa{
\mt{H}_{\ka} = -\tfrac{\la}{2}
\sum_{i \neq j} p_{i-j} \big( \si^+_i \si^-_{j} + \si^-_i \si^+_{j} + \tfrac12\si^z_i \si^z_{j} \big) , \label{HamIno2}
}
so that $\lim\limits_{\ka\to\infty} \mt{H}_\ka = \mt{H}_\xxx$ when $L\to \infty$ .
It is worth noting that the Haldane-Shastry model on the circle also exhibits a $\mc{Y}(\mf{sl}_2)$ Yangian symmetry \cite{LongYangian} and thus the folding could be applied to its Hamiltonian to obtain integrable boundary long-range Hamiltonians on a segment. Symmetries of the latter model using the transfer matrix techniques were studied in \cite{BPS}.


\paragraph{Magnetic boundary}

Our goal is to construct a long-range analogue of the Hamiltonian \eqref{HxxxM}, which exhibits a $\mc{Y}^+(\mf{sl}_2)$ twisted Yangian symmetry. We will achieve this by applying the folding $f$ to $\mt{H}_{\kappa}$ and setting folding constants to the same values as for the semi-infinite XXX spin chain with magnetic boundary, {\it i.e.} those given by \eqref{k1} and
\eq{
k^{zz}=1 ,\qu k^{-+} - k^{+-} = \tfrac{4\mu}{\la} , \qu k^{\pm z}= k^{z\pm} = \pm \tfrac{\la}{\mu}. \label{k1a}
}
Introduce the operators
\eqa{
\mt{H}_{\ka}^{\lo} =& \tfrac{\la}{2}\sum_{\substack{ i\neq j\\i,j \leq 0}}p_{i+j-1}\big(\sigma^+_i\sigma^-_j + \sigma^-_i\sigma^+_j + \tfrac12\sigma^z_i \sigma^z_j\big), \label{Hk}
\\
\mt{M}_{\ka}^{\mu}=& -\tfrac{\la}{2}\sum_{\substack{ i\neq j\\i,j \leq 0}}p_{i+j-1}\si^z_i \si^z_j+\mu \sum_{i \leq 0}p_{2i-1}\si_i^z , \label{Mk} \\[-2.25em]\nonumber
}
satisfying $\lim\limits_{\ka\to\infty} \mt{H}_{\ka}^\lo = 0$ and $\lim\limits_{\ka\to\infty} \mt{M}^\mu_\ka = \mu \si^z_0$.
Then similar computations to those in \eqref{fH1} yield
\eq{
f(\mt{H}_{\kappa}) = 2(\mt{H}_{\ka}^{\mi} +\mt{H}_{\ka}^\lo+ \mt{M}_\ka^\mu) + \tfrac{\la}{2}(1+k^{+-}+k^{-+})\sum_{i \leq 0}p_{2i-1} . \label{fHk1}
}
Let us explain the meaning of operators listed above: $\mt{H}_{\ka}^{\mi}$ is the Hamiltonian \eqref{HamIno2} restricted to a half-line, $\mt{H}_{\ka}^{\lo}$ is the open boundary operator describing the long-range interaction between the sites labelled $i$ and $j$ via the boundary, {\it i.e.} at the distance $i+j-1$, and $\mt{M}_{\ka}^{\mu}$ is the the long-range magnetic boundary operator; for both $\mt{H}^\lo_\ka$ and $\mt{M}_{\ka}^{\mu}$ their numerical values decay exponentially moving away from the boundary. Hence, by neglecting the constant term in \eqref{fHk1}, we conclude that
\eq{
\mt{H}_{\ka}^{\mu} := \mt{H}_{\ka}^{\mi}+ \mt{H}_{\ka}^\lo+ \mt{M}_\ka^\mu
}
is the Hamiltonian of the open Inozemtsev hyperbolic spin chain with a magnetic boundary. It will be shown below that it is integrable, {\it i.e.} exhibits a Yangian symmetry, only if $\mu=\pm\la$.

We already know that $f(\mt{E}^a_0)=2 \delta_{az}(\mt{E}^a_0)^\mi$, which remains the only Lie symmetry of $\mt{H}_{\ka}^{\mu}$. Under $f$, the operators \eqref{YangIno} are mapped to
\eqn{
f &(\mt{E}_{\ka,1}^{\pm}) = \\&= \pm \tfrac{\la}{2} \Bigg(\sum_{i,j\leq 0} k^{\pm 0}k^{z0}w_{i-j}\si^{\pm}_i \si^z_j+\sum_{i,j>0}k^{0z}k^{0 \pm}w_{i-j}\si_{1-i}^{\pm} \si_{1-j}^z
\\
& \qq\qu + \sum_{\substack{i \leq 0, j>0 \\ i \neq 1-j}}k^{0z}w_{i-j}\si_i^{\pm} \si_{1-j}^z+ \sum_{\substack{j \leq 0, i>0 \\ j \neq 1-i}}k^{0 \pm}w_{i-j}\si_{1-i}^{\pm} \si_j^z
\\
&\qq\qu + \sum_{i \leq 0}k^{\pm z}w_{2i-1}\si_i^{\pm} \si_{i}^z+ \sum_{i \leq 0}k^{z \pm}w_{1-2i}\si_{i}^z \si_i^{\pm}\Bigg)
\\
&= 2\mt{E}_1^{\pm}
\pm \la \sum_{\substack{i\ne j \\i,j \leq 0}} w_{i+j-1} \si_i^{\pm} \si_j^z-\tfrac{\la}{2} (k^{z \pm}+k^{\pm z}) \sum_{i \leq 0}\ w_{2i-1} \si^{\pm}_i
}
and
\eqn{
f&(\mt{E}_{\ka,1}^{z}) = \\ &= \la \Bigg(\sum_{i,j\leq 0} k^{+ 0}k^{-0}w_{i-j}\si^{+}_i \si^-_j+\sum_{i,j>0}k^{0-}k^{0 +}w_{i-j}\si_{1-i}^{+} \si_{1-j}^-
\\
& \qq\qu + \sum_{\substack{i \leq 0, j>0 \\ i \neq 1-j}}k^{0-}w_{i-j}\si_i^{+} \si_{1-j}^-+ \sum_{\substack{j \leq 0, i>0 \\ j \neq 1-i}}k^{0 +}w_{i-j}\si_{1-i}^{+} \si_j^-
\\
&\qq\qu + \sum_{i \leq 0}k^{+-}w_{2i-1}\si_i^{+} \si_{i}^-+ \sum_{i \leq 0}k^{-+}w_{1-2i}\si_{i}^- \si_i^{+}\Bigg)
\\
&=-\tfrac{\la}{2} (k^{+-}+k^{-+}) \sum_{i\leq 0} w_{2i-1} \si^z_i - \tfrac{\la}{2} (k^{+-}-k^{-+}) \sum_{i\leq 0} w_{2i-1} .
}
The folded operators $f(\mt{E}_{\ka,1}^{\pm})$ satisfy the defining relations of the $\mc{Y}^+(\mf{sl}_2)$ twisted Yangian for any $L>0$ provided \eqref{k1a} holds. It remains to verify if they are symmetries of $\mt{H}_\ka^\mu$. It is straightforward to see that $[\mt{H}^\mu_\ka,f(\mt{E}^{z}_{\ka,1})]=0$ . By computing the commutator $[\mt{H}^\mu_\ka,f(\mt{E}^{\pm}_{\ka,1})]$ we find that it equals to zero in the $L\to\infty$ limit only and provided \eqref{k1} and the following constraints hold
\eq{
k^{+-} = - k^{-+} = \pm 2, \qu k^{z-} = -k^{z+} = \mp\tfrac{1}{2} . \label{k1b}
}
In other words, $f(\mt{E}^{a}_{\ka,1})$ are symmetries of $\mt{H}^\mu_\ka$ only if $\mu=\pm\la$ thus implying the aforementioned integrability condition for the long-range Hamiltonian $\mt{H}_\ka^\mu$. In particular, its Yangian symmetries are
\eq{
\mt{X}^{\pm}_\ka = (\mt{E}_{\ka,1}^{\pm})^\mi\pm\tfrac{\la}{2}\sum_{\substack{i\ne j\\ i,j \leq 0}} w_{i+j-1}\si_i^{\pm} \sigma_j^z \pm \tfrac{\la^2}{2 \mu}\sum_{i \leq 0}  w_{2i-1} \si^{\pm}_i,
}
with $\mu=\la$ or $\mu=-\la$ the two cases being related to each other via the Lie algebra automorphism $\theta : \si^{\pm}\mapsto\si^\mp$, $\si^z\mapsto -\si^z$. This automorphism leaves the Hamiltonian $\mt{H}_\ka$ (and $\mt{H}_\ka^\mi$, $\mt{H}_\ka^\lo$) invariant, but maps $\mt{H}_{\ka}^{\la}$ to $\mt{H}_{\ka}^{-\la}$ and $\mt{X}^{\pm}_{\ka}$ to $\mt{X}^{\mp}_{\ka}$.

We conclude this section with two remarks. First, by applying the same folding procedure to the symmetries $\mt{E}^{\prime \pm}_{\ka,1}$ and $\mt{E}^{\prime\prime \pm}_{\ka,1}$ of $\mt{H}_\ka$ we obtain operators $\mt{X}^{\prime \pm}_\ka = f(\mt{E}^{\prime \pm}_{\ka,1})$ and $\mt{X}^{\prime\prime \pm}_\ka = f(\mt{E}^{\prime\prime \pm}_{\ka,1})$ that are symmetries of $\mt{H}_\ka^\mu$ provided \eqref{k1b} holds. They are long-range analogues of the operators \eqref{half-Yang2}. Second, assuming that $\mu\in\C$ is arbitrary and taking the $\ka\to\infty$ limit, operators $\mt{X}^{\pm}_\ka$ and $\mt{X}^{\prime \pm}_\ka$, $\mt{X}^{\prime\prime \pm}_\ka$ specialize to their nearest-neighbour counterparts given in \eqref{Yang2} and \eqref{half-Yang2}.


\paragraph{Open boundary}

We want to construct a long-range analogue of the Hamiltonian \eqref{Hxxx0}, which exhibits a $\mc{Y}^-(\mf{sl}_2)$ twisted Yangian symmetry. We will achieve this by applying the folding $f$ with $k^{ab}=1$ to $\mt{H}_{\kappa}$. In particular, we find that
\eq{
f(\mt{H}_{\kappa}) = 2(\mt{H}_{\ka}^{\mi} + \mt{H}_\ka^\lo ) - \tfrac{3}{2}\la \sum_{i\le 0}p_{2i-1} ,
}
which, after dropping the constant term, is the open boundary Hamiltonian as expected from \eqref{fHk1}.

To obtain Yangian symmetries of the long-range open boundary model we need to fold the long-range analogues of the operators \eqref{Goodlev2}:
\eqn{
\wt{\mt{E}}^+_{\ka,2} &= \mt{E}^+_{\ka,2} + \tfrac13 \big([\mt{E}^{\prime z}_{\ka,2}, \mt{E}^{\prime\prime +}_{\ka,2}] + [\mt{E}^{\prime\prime z}_{\ka,2}, \mt{E}^{\prime +}_{\ka,2}] \big) + \tfrac{\la^2}{3} \big( \mt{E}^+_0 \mt{E}^-_0\mt{E}^+_0 - \tfrac94 \mt{E}_0^+ \big) ,
\\	
\wt{\mt{E}}^-_{\ka,2} &= \mt{E}^-_{\ka,2} - \tfrac13 \big([\mt{E}^{\prime z}_{\ka,2}, \mt{E}^{\prime -}_{\ka,2}] + [\mt{E}^{\prime\prime z}_{\ka,2}, \mt{E}^{\prime\prime -}_{\ka,2}] \big) + \tfrac{\la^2}{3} \big( \mt{E}^-_0 \mt{E}^+_0\mt{E}^-_0 - \tfrac94 \mt{E}_0^- \big) ,
\\	
\wt{\mt{E}}^z_{\ka,2} &= \mt{E}^z_{\ka,2} + \tfrac23 \big( [\mt{E}^{\prime +}_{\ka,2}, \mt{E}^{\prime -}_{\ka,2}] + [\mt{E}^{\prime\prime +}_{\ka,2}, \mt{E}^{\prime\prime -}_{\ka,2}] \big) + \tfrac{\la^2}{6} \big( (\mt{E}^z_0)^3 - \tfrac72 \mt{E}_0^z \big) .
}
By doing so we find
\eqn{
f(\wt{\mt{E}}^+_{\ka,2}) &= \tfrac{16}{3}(\mt{E}^{\pm}_{\ka,2})^{\mi} \\ &+\tfrac{\la^2}{3} \sum_{i,j,k} a_{ijk}\big( \si^z_i \si^z_j \si^{\pm}_k +4\si^+_i \si^-_j \si^{\pm}_k \big)+\tfrac{2\la^2}{3}\sum_{i,j} b_{ij} \si^{\pm}_i , \\
f(\wt{\mt{E}}^z_{\ka,2}) &= \tfrac{16}{3} (\mt{E}^{z}_{\ka,2})^{\mi} \\ &+\tfrac{\la^2}{3} \sum_{i,j,k} a_{ijk}\big( \si^z_i \si^z_j \si^z_k+4\si^+_i \si^-_j \si^z_k \big)+\tfrac{2\la^2}{3}\sum_{i,j} b_{ij} \si^{z}_i ,
}
where
\eqn{
a_{ijk}&=2-w_{i-j}(w_{j-k}+w_{i+k-1}-w_{i-k}-w_{j+k-1})\\
& -w_{i+j-1}(w_{i-k}+w_{j-k}+w_{i+k-1}+w_{j+k-1}), \\
b_{ij}&=5+w_{i-j}^2-\frac{1}{4}w_{1-2i}^2-w_{i+j-1}(w_{i+j-1}-4 w_{1-2j})\\
& -2 w_{i-j}(w_{i+j-1}+2 w_{1-2j}) .
}
The operators $\mt{G}^a_{\ka} = \tfrac38 f(\wt{\mt{E}}^a_{\ka,2})$ together with $(\mt{E}_0^a)^\mi$ satisfy the defining relations of the $\mc{Y}^-(\mf{sl}_2)$ twisted Yangian and commute with the Hamiltonian $\mt{H}_\ka^0 = \mt{H}_\ka^\mi + \mt{H}^{\lo}_\ka$, up to the terms at infinity.

We also have that $\lim\limits_{\ka\to\infty} \mt{G}^a_\ka = \mt{G}^a$ and there are a number of additional symmetries of $\mt{H}_\ka^0$ that are obtained by folding quadratic combinations of the symmetries $\mt{E}_{\ka,1}^{\prime a}$ and $\mt{E}_{\ka,1}^{\prime\prime a}$ of $\mt{H}_\ka$.


\paragraph{Double-row chain with a diagonal boundary}

Let us now focus on the model consisting of two uncoupled Inozemtsev hyperbolic spin chains described by the Hamiltonian
\eqa{
\mt{H}_{\kappa}^{\circ\bullet} = -\tfrac{\la}{2} \sum_{\al=\circ,\bullet}  \sum_{i\neq j} p_{i-j}\big(\si^+_{i\al} \si^-_{j,\al} + \si^-_{i\al} \si^+_{j\al} + \tfrac12 \si^z_{i\al} \si^z_{j\al}\big). 
\label{Ino2}
}
As the double-row XXX model this model exhibits a $\mc{Y}(\mf{so}_4)$ Yangian symmetry generated by the Lie operators $\mt{E}^{a}_{0\al}$ and the double-row analogues $\mt{E}^a_{\ka1\al}$ of the ones defined in \eqref{YangIno}.

We use the folding $\overline f$ with $k^{ab}=1$ to obtain an integrable long-range analogue of the Hamiltonian \eqref{HxxxD} exhibiting a $\mc{Y}^{{\Delta}}(\mf{sl}_2)$ twisted Yangian symmetry.

Introduce the operator
\eq{
\!\!\mt{D}_\ka = -\tfrac{\la}{2}\!\!\!\!\sum_{\substack{\al \neq \beta\\ \al,\beta=\circ,\bullet}} \!\!\!\sum_{i,j \leq 0} \!\! p_{i+j-1}\big( \si^+_{i\al} \si^-_{j\beta} + \si^-_{i\al} \si^+_{j\beta} + \tfrac12\si^z_{i\al} \si^z_{j\beta}\big).
}
Proceeding in a similar way as for the double-row XXX model we have that
\eqn{
\overline f(\mt{H}_{\ka}^{\circ\bullet})&=2\big( (\mt{H}_{\kappa}^{\circ\bullet})^\mi + \mt{D}_\ka \big).
}
The operator $\mt{D}_\ka$ is the long-range diagonal boundary operator for the semi-infinite long-range double-row model; it can also be viewed as a double-row analogue of the open boundary operator $\mt{H}^\lo_\ka$. In the $\ka\to\infty$ limit $\mt{D}_\ka$ specializes to boundary term in \eqref{HxxxD2}.

Next we fold the the long-range analogues of the operators \eqref{so4}. Similarly as before we have that $\overline{f}(\mt{A}_{\ka,1}^a ) = 0$ and
\eqn{
\overline f(\mt{B}_{\ka,1}^{\pm}) &= 2(\mt{B}_{\ka,1}^{\pm})^\mi \pm \la \sum_{i,j \leq 0} w_{i+j-1}(\si^{\pm}_{i,\circ} \si^z_{j,\bullet}-\si^{\pm}_{j,\bullet}\si^z_{i,\circ}) ,\\
\overline f(\mt{B}_{\ka,1}^{z}) &= 2(\mt{B}_{\ka,1}^{z})^\mi - 2 \la \sum_{i,j \leq 0} w_{i+j-1}(\si^+_{i,\circ} \si^-_{j,\bullet}-\si^+_{j,\bullet}\si^-_{i,\circ}) .
}
The operators $\mt{Y}_{\ka,1}^a = \frac12 \overline f(\mt{B}_{\ka,1}^a)$ together with $\mt{A}^a_0$ satisfy the defining relations of the $\mc{Y}^{{\Delta}}(\mf{sl}_2)$ twisted Yangian and commute with the Hamiltonian $\mt{H}_\ka^\Delta = (\mt{H}_{\xxx}^{\circ\bullet})^\mi + \mt{D}_\ka$ up to the terms at infinity. In the $\ka\to\infty$ limit $\mt{Y}_{\ka,1}^a$ specialize to those given in \eqref{Yang3}.

We also remark that there exists a number of symmetries of $\mt{H}_\ka^\Delta$ that are obtained by folding the double-row analogues of the operators $\mt{E}^{\prime a}_{\ka,1}$ and $\mt{E}^{\prime\prime a}_{\ka,1}$. These additional symmetries also specialize to those of the double-row XXX model.


\section{Conclusions and Outlook} \label{sec:5}

In this letter we have presented a method for constructing integrable boundaries for $\mf{sl}_2$-symmetric spin chains and their doublings without relying on the boundary Yang Baxter equation. This method, which we refer to as ``folding'', consists in a map denoted by $f$ (and $\overline f$ in the doubled case) which sends the operators of a model defined on the infinite line to those on the half-line.

More precisely, given a Hamiltonian $\mt{H}$ of an infinite spin chain and a family of operators $\{\mt{Q}_\al\}_{\al\in I}$ indexed by a set $I$ and commuting with the Hamiltonian, $[\mt{H},\mt{Q}_\al]=0$ for all $\al\in I$, the folding identifies the positive half-line with the negative half-line in such a way that, for a suitable choice of the ``folding constants'', the folded Hamiltonian $f(\mt{H})$, which now describes a semi-infinite spin chain, commutes with the folded operators, namely $[f(\mt{H}),f(\mt{Q_\al})]=0$ for all $\al\in I$.

The choice of the folding constants is dictated by the symmetry properties of the Hamiltonian $\mt{H}$ and the would-be symmetries of the folded Hamiltonian $f(\mt{H})$. Integrability is then ensured by the existence of an infinite number of conserved quantities, {\em i.e.\ }operators commuting with the Hamiltonian and satisfying the defining relations of an infinite-dimensional algebra \cite{MK1}.
In the case when the Hamiltonian exhibits a $\mc{Y}(\mf{sl}_2)$ Yangian symmetry there are three non-equivalent boundary integrable models that can be obtained: a spin chain with a magnetic boundary, a spin chain with an open boundary, and a double-row model with a diagonal boundary. These models exhibit $\mc{Y}^+(\mf{sl}_2)$, $\mc{Y}^-(\mf{sl}_2)$ and $\mc{Y}^{{\Delta}}(\mf{sl}_2)$ twisted Yangian symmetries, respectively. For the Heisenberg XXX spin chain the corresponding models are well-studied. However, this is not the case for the Inozemtsev hyperbolic spin chain. Integrable boundary Hamiltonians for the latter were constructed in \cite{BPS, XW} using the Dunkl operators and, although similar in form, the results obtained in {\it loc.~cit.} differ from ours. It remains to be shown whether folding can yield those boundary Hamiltonians and if they exhibit any Yangian symmetries. This is natural to expect, since Hamiltonians of such type were shown to obey infinite dimensional symmetries \cite{CC1,CC2}.

The method presented in this letter can be easily applied to any integrable spin chains. Let $\mf{g}$ be any simple Lie algebra of rank$(\mf{g})\ge2$ and let $\mt{H}_\mf{g}$ be a spin chain Hamiltonian exhibiting $\mc{Y}(\mf{g})$ Yangian symmetry. Let $\theta : \mf{g} \to \mf{g}$ be an involutive automorphism of $\mf{g}$. Denote by $\mf{h}=\mf{g}^\theta$ the $\theta$-fixed subalgebra, so that $(\mf{g},\mf{h})$ is a symmetric pair. For such a pair there exists an infinite dimensional algebra, the $\mc{Y}(\mf{g},\mf{h})$ twisted Yangian, which is a coideal subalgebra of $\mc{Y}(\mf{g})$ \cite{MK2,BR}, and there exists a boundary-integrable spin chain exhibiting such a symmetry, which can be constructed using the folding method presented in this letter. While this might be rather straightforward for spin chains with nearest neighbor interactions only, since the boundary term for such models in many cases is a symmetry breaking term exhibiting $\mf{h}$-symmetry only, this is no longer true for the long-range spin chains, as we have shown in this letter. Moreover, obtaining long-range Hamiltonians using the techniques of the inverse scattering method is a rather challenging task, as it was shown in \cite{BPS, XW}, thus the ``bottom-up'' approach provides a short-cut for constructing such models.

It is important to note that the folding $f$ is generally not an algebra homomorphism. It can only be so if $\mf{h}$ is a commutative subalgebra of $\mf{g}$. The only symmetric pair satisfying this requirement is $(\mf{g},\mf{h})=(\mf{sl}_2,\mf{gl}_1)$, which we have studied in this letter. In all other cases the map $f$ is effectively a projector.

Another important thing to note is that folding is only a good method of constructing boundary-integrable models if it is defined over a link. If we instead fold at a site, symmetry arguments force the folding constants associated with that site to be zero, thus effectively turning folding over a site into folding over a link.

We finish by noting that a very interesting subject for our folding method would be the Hubbard model \cite{Hubbard} and its long-range analogue \cite{IG}. It would be interesting to see if one can gain further insight into the unusual structure of the Hubbard model's known integrable boundaries \cite{Gomez,SW,GN} and perhaps obtain new ones.

\smallskip

\noindent {\it Acknowledgements.} V.R.~was partially supported by the UK EPSRC under the grant EP/K031805/1. The authors would also like to thank R. Frassek, R. Klabbers, J. Lamers and I. Szecsenyi for useful discussions.


\appendix


\newcounter{AX}[section]

\section{}  \label{App:A}

We briefly recall the necessary details of the Yangian $\mc{Y}(\mf{sl}_2)$ and twisted Yangians $\mc{Y}^+(\mf{sl}_2)$ and $\mc{Y}^-(\mf{sl}_2)$ (adhering to \cite[Sec.~5.1]{GRW}; see also \cite{Olshanski}), and the diagonal twisted Yangian $\mc{Y}^{\Delta}(\mf{sl}_2)$ (called ``achiral'' in \cite{MR}). 
All the representations described below are related to the Heisenberg XXX spin chain. In case of the Inozemtsev hyperbolicspin chain the operators $\mt{E}^a_1$, $\mt{X}^\pm$, {\it etc.} are with their long-range counterparts; that is, $\mt{E}^a_{\ka,1}$, $\mt{X}^\pm_\ka$, {\it etc.}


\phase{} \label{A1} 
Let $\la\in\C^\times$. The Yangian $\mc{Y}(\mf{sl}_2)$ is generated by the elements $x^\pm$, $h$ and $J(x^\pm)$, $J(h)$ satisfying
\spl{
& [h,x^\pm]=\pm2x^\pm,\qu [x^+,x^-] = h, \\
& [J(h),x^\pm]=[h,J(x^\pm)]=\pm2J(x^\pm), \qu [J(x^\pm),x^\mp] = \pm J(h),\\
& [J(h),[J(x^+),J(x^-)]] = \la^2\big( J(x^-) x^+ - x^- J(x^+)\big) h. \label{Y2}
}
The representation on the infinite spin chain is given by the map $\rho_\infty : \mc{Y}(\mf{sl}_2) \to \Si_\infty$  defined by
\eq{
x^\pm \mapsto \mt{E}^\pm_0, \qu h \mapsto \mt{E}^z_0, \qu
J(x^\pm) \mapsto \mt{E}^\pm_1, \qu J(h) \mapsto \mt{E}^z_0. \label{Rep:Y2}
}


\phase{} \label{A2} 
Let $c\in\C$. The one-parameter twisted Yangian $\mc{Y}^+(\mf{sl}_2)$ for the symmetric pair $(\mf{sl}_2,\mf{gl}_1)$ is generated by the elements $k$ and $B(x^\pm)$ satisfying
\eqa{
& [k,B(x^\pm)]=\pm2B(x^\pm), \label{Y2+} \\
& [B(x^\pm),[B(x^\pm),[B(x^\mp),B(x^\pm)]]] = 12 \la^2 B(x^\pm) (k+c) B(x^\pm) . \nn 
}
Let $\al^\pm\in\C$ be such that $\al^+ - \al^-=2c$. The embedding \linebreak $\varphi^+ : \mc{Y}^+(\mf{sl}_2) \hookrightarrow \mc{Y}(\mf{sl}_2)$ of algebras is given by
\eqa{
k \mapsto h , \qu B(x^\pm) \mapsto J(x^\pm) \pm \tfrac{\la}2 x^\pm h + \la \al^\pm x^\pm.
}
Set $c=-\la/\mu$. The representation on the half-infinite spin chain is given by the map $\rho^+_\infty : \mc{Y}^+(\mf{sl}_2) \to \Si^\mi_\infty$ defined by
\eq{
h \mapsto (\mt{E}^z_0)^\mi, \qu
B(x^\pm) \mapsto \mt{X}^\pm. \label{Rep:Y2+}
}


\phase{} \label{A3} 
The twisted Yangian $\mc{Y}^-(\mf{sl}_2)$ for the trivial symmetric pair $(\mf{sl}_2,\mf{sl}_2)$ is generated by the elements $x^\pm$, $h$ and $G(x^\pm)$, $G(h)$ satisfying
\eqa{
& [h,x^\pm]=\pm2x^\pm,\qu [x^+,x^-] = h,  \el
& [G(h),x^\pm]=[h,G(x^\pm)]=\pm2J(x^\pm),\qu [G(x^\pm),x^\mp] = \pm G(h), \el
& [G(h),[G(x^+),G(x^-)]] = 4\la^2 \big( \{x^+,G(x^-), G(h) \} \el & \hspace{4.1cm}- \{x^-,G(x^+), G(h) \} \big) , \label{Y2-} 
}
where $\{x_1,x_2,x_3\} = \frac16 \sum_{p \in S_3} x_{p(1)} x_{p(2)} x_{i(3)}$ is the normalized total symmetrizer. The embedding $\varphi_{-} : \mc{Y}^-(\mf{sl}_2) \hookrightarrow \mc{Y}(\mf{sl}_2)$ of algebras is given by
\eqa{
x \mapsto x, \qu G(x) \mapsto [J(x'),J(x'')] + \tfrac{\la}4 [J(x),C] - \tfrac{\la^2}{4} x
}
for all triples $(x,x',x'')\in \{(h,x^+,x^-),(e,\half h,x^+)$, $(f,x^-,\half h)\}$ and all $x\in\{x^\pm,h\}$; here $C=x^+x^- + x^-x^+ + \half h^2$ is the quadratic Casimir.
The representation on the half-infinite spin chain is given by the map $\rho^-_\infty : \mc{Y}^-(\mf{sl}_2) \to \Si^\mi_\infty$ defined by
\eq{
x^\pm \mapsto (\mt{E}^\pm_0)^\mi, \qu h \mapsto (\mt{E}^z_0)^\mi, \qu
G(x^\pm) \mapsto \mt{G}^\pm, \qu G(h) \mapsto \mt{G}^z. \label{Rep:Y2-}
}


\phase{} \label{A4}
The diagonal twisted Yangian $\mc{Y}^{{\Delta}}(\mf{sl}_2)$ the symmetric pair $(\mf{sl}_2\op\mf{sl}_2,\mf{sl}_2)$ is generated by the elements $h, x^\pm$ satisfying the usual $\mf{sl}_2$ Lie algebra relations and $D(h)$, $D(x^\pm)$ satisfying
\eqa{
& [D(h),x^\pm]=[h,D(x^\pm)]=\pm2D(x^\pm), \qu [D(x^\pm),x^\mp] = \pm D(h), \nonumber\\
& [D(h),[D(x^+),D(x^-)]] = \la^2( D(x^-) x^+ - x^- D(x^+)) h.  \label{Y2D}
}
The embedding $\varphi^\Delta : \mc{Y}^{\Delta}(\mf{sl}_2) \hookrightarrow \mc{Y}^\circ(\mf{sl}_2) \ot \mc{Y}^\bullet(\mf{sl}_2)$ of algebras, where $^{\circ}$ and $^{\bullet}$ are used to distinguish two copies of $\mc{Y}(\mf{sl}_2)$,  is given by
\spl{
& h \mapsto h_\circ + h_\bullet , \qu x^\pm \mapsto x^+_\circ + x^+_\bullet,
\\
& D(h) \mapsto J(h_\circ) - J(h_\bullet) - \la \big((x^+_\circ - x^+_\bullet)(x^-_\circ + x^-_\bullet)  \\
& \hspace{4cm} - (x^+_\circ + x^+_\bullet)(x^-_\circ - x^-_\bullet)\big) ,
\\
&D(x^\pm) \mapsto J(x^\pm_\circ) - J(x^\pm_\bullet) \pm \tfrac\la2 \big((x^\pm_\circ - x^\pm_\bullet)(h_\circ + h_\bullet) \\
& \hspace{4.5cm} - (x^\pm_\circ + x^\pm_\bullet)(h_\circ - h_\bullet)\big).
}
The representation on the double-row half-infinite spin chain is given by the map $\rho^\Delta_\infty : \mc{Y}^{{\Delta}}(\mf{sl}_2) \to \overline{\Si}{}^\mi_\infty$ defined by
\eq{
h \mapsto (\mt{A}^z_{0})^\mi, \qu\!\!
x^\pm \mapsto (\mt{A}^\pm_{0})^\mi, \qu\!\!
D(h) \mapsto \mt{Y}^z, \qu\!\!
D(x^\pm) \mapsto \mt{Y}^\pm. \label{Rep:Y2D}
}


\end{document}